# Very Strong Superconducting Proximity Effects in PbS Semiconductor Nanowires

*Bum-Kyu Kim[1], Hong-Seok Kim[1], Yiming Yang[2], Xingyue Peng[2], Dong Yu[2], Yong-Joo Doh[1,*]*

[1]Department of Physics and Photon Science, Gwangju Institute of Science and Technology (GIST), Gwangju, 61005, Korea

[2]Department of Physics, University of California, Davis, CA 95616, USA

**ABSTRACT**

We report the fabrication of strongly coupled nanohybrid superconducting junctions using PbS semiconductor nanowires and $Pb_{0.5}In_{0.5}$ superconducting electrodes. The maximum supercurrent in the junction reaches up to ~15 μA at 0.3 K, which is the highest value ever observed in semiconductor-nanowire-based superconducting junctions. The observation of microwave-induced constant voltage steps confirms the existence of genuine Josephson coupling through the nanowire. Monotonic suppression of the critical current under an external magnetic field is also in good agreement with the narrow junction model. The temperature-dependent stochastic distribution of the switching current exhibits a crossover from phase diffusion to a thermal activation process as the temperature decreases. These strongly coupled nanohybrid superconducting junctions would be advantageous to the development of gate-tunable superconducting quantum information devices.

When a semiconductor nanowire (NW) is contacted with two superconducting electrodes, a supercurrent can flow through the nonsuperconducting nanowire because of the superconducting proximity effect.[1] The NW-based superconducting junctions,[2, 3] or nanohybrid Josephson junctions, acquire an advantage of a gate-tunable supercurrent flow by controlling the carrier density in the NW, resulting in a supercurrent-based field-effect transistor.[4] When the supercurrent is combined with phase-coherent quantum electronic transport in the NW, semiconductor NWs provide a useful platform to develop novel quantum electronic devices, such as gate-tunable superconducting quantum interference devices,[5, 6] Cooper pair splitters,[7] and quantum electron pumps,[8] and to explore Majorana fermions,[9, 10] and gate-tunable superconducting qubits.[11, 12]

So far, most of the NW-based superconducting junctions have been made of Al superconducting electrodes with a very low transition temperature at $T_C \sim 1.2$ K. The ultralow transition temperature and very small superconducting gap energy of Al, $\Delta_{Al} \sim 0.15$ meV, result in a relatively small critical current $I_C$ ($\leq 0.1$ μA), which is defined as the maximum supercurrent. Recently, a very short ($L \sim 30$ nm) channel device[13] of InAs NW contacted with Al electrodes has exhibited a maximum supercurrent up to $I_C \sim 0.8$ μA. Because the Josephson coupling energy $E_J = \hbar I_C/2e$, where $\hbar$ is Planck constant $h$ divided by $2\pi$ and $e$ is the elementary charge, must exceed thermal fluctuations for the observation of nonzero supercurrent, a large critical current would be essential for a wide variety of applications of the NW-based superconducting devices. Although other superconducting electrodes with higher transition temperatures, such as V ($T_C \sim 5.0$ K),[14] Pb ($T_C \sim 7.2$ K),[15] and Nb ($T_C \sim 9.2$ K),[16] have been used to increase the critical current of NW-based superconducting junctions, the improvements ($I_C < 1$ μA) were not significant.

In this work, we demonstrate very strong superconducting proximity effects between a PbS semiconductor NW and PbIn superconducting electrodes. The maximum supercurrent reaches up to $I_C \sim 15$ μA at $T = 0.3$ K, which is, to the best of our knowledge, the highest value for semiconductor-NW-based superconducting junctions. Moreover, the $I_C R_N$ product, which is a figure of merit for a Josephson junction with a normal-state resistance $R_N$, is larger than the superconducting gap energy of PbIn, resulting in $eI_C R_N / \Delta_{PbIn} = 1.05$, which is the largest value reported so far for the semiconducto-NW-based superconducting junctions. We also examined the superconducting proximity effects in presence of the microwave and magnetic fields, which are consistent with the theoretical expectations. Furthermore, our measurements of the switching current distribution from the superconducting to resistive branches reveal that the phase diffusion and thermal activation processes are responsible for the stochastic switching behavior, which is dependent on temperature.

## RESULTS AND DISCUSSION

The PbS NWs were synthesized using the chemical vapor deposition method in a tube furnace, while the doping of the NWs was modulated by varying the weight ratio of lead chloride and sulfur (see Methods). Figure 1a shows the scanning electron microscopy (SEM) image of the as-grown PbS NWs. The details of NW growth[17] and device fabrication[6] have been reported elsewhere. A representative SEM image of the PbS NW device is displayed in Fig. 1b, where two neighboring superconducting PbIn electrodes were used to apply the bias current (from I+ to I−) and to measure the voltage difference (between V+ and V−). The diameter $w$ of the PbS NW and the distance $L$ between two superconducting electrodes are found to be 130–200 nm and 180–220 nm, respectively, while the normal-state resistance $R_N$ of the device ranges from 20–130 $\Omega$ (see Table S1). The electrical transport properties of the NW devices were measured using a closed-cycle $^3$He refrigerator (Cryogenic Ltd.) down to the base temperature of 0.3 K. For the low-noise measurements, two-stage RC filters (cutoff frequency ~30 kHz) and $\pi$ filters were connected in series to the measurement leads.[18]

The PbIn superconducting electrode exhibits a superconducting transition below $T_{C,SC}$ = 7.0 K upon lowering the temperature, while the supercurrent can flow through the NW junction below $T_{C,JJ}$ = 3.5 K for device **D3**, as shown in Fig. 1c. The highest temperature for the observation of the supercurrent was found to be $T$ = 5.2 K in PbS NW-based Josephson junctions.[6] Figure 1d shows the current–voltage ($I$–$V$) characteristic curves of devices **D1** and **D2**, displaying hysteresis depending on the sweep direction of the bias current. The switching from the superconductive to dissipative branches occurs at a critical current $I_C$, while the reversed switching occurs at a return current $I_R$. The hysteresis can be understood in terms of the presence of an effective capacitance[19, 20] in the junction or the quasiparticle heating effect.[21] Here, we note that device **D1** exhibits a maximum switching current $I_C$ ~ 15 μA, corresponding to a supercurrent density $J_C$ ~ 1.1 × 10$^5$ A/cm$^2$. To the best of our knowledge, these are the highest $I_C$ and $J_C$ values recorded for the semiconductor-NW-based superconducting junctions,[13-16, 22] which is rather close to those observed for Au-NW-based junctions.[20] For device **D2**, the $eI_CR_N$ product is larger than $\Delta_{PbIn}$, where $\Delta_{PbIn}$ = 0.81 meV is the superconducting gap energy (see discussion below). Table 1 shows a summary of the junction properties for a quantitative comparison.

The differential conductance curve as a function of voltage, $dI/dV(V)$, is shown in Fig. 1e for device **D1** at $T = 0.3$ K. The overshoot of $dI/dV$ near zero voltage is caused by the supercurrent branch, while the $dI/dV$ peaks are attributed to the existence of multiple Andreev reflections[23] (MARs) at the interfaces between the PbIn superconducting electrodes and the PbS semiconductor NW. When an electron coming from the semiconductor is incident upon the highly transparent interface, it can be retro-reflected as a phase-conjugated hole (i.e., the lack of an electron below the Fermi energy) while leaving a Cooper pair in the superconductor, known as the Andreev reflection.[1] When the Andreev reflections occur successively at two superconductor-semiconductor interfaces on the opposite side of the junction, the MARs result in conductance enhancements (or $dI/dV$ peaks) occurring at $V_n^* = 2\Delta/ne$, where $n$ is an integer and $\Delta$ is the superconducting gap energy. It is evidently shown that the $dI/dV$ peaks occur at $V_1^* =$ 1.62 mV and $V_2^* = 0.81$ mV, indicating $\Delta_{PbIn} = 0.81$ meV. The temperature dependence of the $dI/dV$ peak positions turns out to be consistent with the Bardeen-Cooper-Schrieffer (BCS) theory of superconductivity[1] (see Fig. S1). Similar features of the subgap structures can be found for different device, implying the transparency at the interface to be $T_{int} = 0.86$, which is deduced from the excess current (see Fig. S2). We note that this value is higher than those obtained from previous superconducting junctions based on InAs semiconductor NWs.[2, 13, 16] The large superconducting gap energy of $\Delta_{PbIn}$ and formation of highly transparent contacts between the PbIn electrodes and the PbS NW are responsible for the very strong Josephson coupling observed in our experiment. An application of the gate voltage $V_G$ can tune the values of $I_C$ and $R_N$, as shown in Fig. 1f. Here, the $I_C(V_G)$ and $R_N(V_G)$ data were taken at $T = 2.5$ K and $T = 10.0$ K, respectively. The increase in $R_N$ with decreasing $V_G$ indicates that the PbS NW has a strong $n$- type character in the experimental range. Correspondingly, $I_C$ decreases with decreasing $V_G$, suggesting a nearly constant $I_C R_N$ product over the whole $V_G$ range in this experiment.

The progressive change in the $I$–$V$ curves with increasing temperature is displayed in Fig. 2a. It is evidently shown that $I_C$ decreases monotonically and the hysteresis becomes reduced at higher temperatures than $T = 1.2$ K. The temperature dependences of $I_C$ and $I_R$ are depicted in Fig. 2b, together with the fitting result for the $I_C(T)$ curve. We fitted $I_C(T)$ to the theoretical expression[24] $eI_C R_N = aE_{Th}[1 - b \exp(-aE_{Th}/3.2k_B T)]$, where $E_{Th} = \hbar D/L^2$ is the Thouless energy, and $a$ and $b$ are fitting parameters. The value of $E_{Th}$ is estimated to be 0.19 meV for the channel length $L = 190$ nm and diffusion coefficient[6] $D = 103$ cm$^2$/s. Thus, the best-fit (the solid-line curve in Fig. 2b) is obtained with $a = 4.6$ and $b = 3$, which are comparable to the parameters in the long and diffusive junction limit[24] ($a = 10.8$ and $b = 1.3$). The difference can be explained by that our PbS NW Josephson junction ($E_{Th}/\Delta_{PbIn} = 0.23$) is in the intermediate regime between the long- ($E_{Th}/\Delta_{PbIn} < 0.01$) and short-junction ($E_{Th}/\Delta_{PbIn} > 1$) limits, which is similar to previously studied nanohybrid superconducting junctions.[18, 20] The superconducting coherence length $\xi = (\hbar D/\Delta_{PbIn})^{1/2}$ is found to be $\xi = 92$ nm, which is about half of the channel length $L$.

The existence of genuine Josephson coupling in the PbS NWs can be confirmed by the microwave response of the NW devices. Under microwave irradiation, the $I$–$V$ curve of the Josephson junction is expected to exhibit quantized voltage plateaus (the so-called Shapiro steps) at $V_n = nhf_{mw}/2e$, where $f_{mw}$ is the frequency of the external microwave.[1] Figure 3a displays the progressive evolution of the $I$–$V$ curve for various values of the microwave power $P$ at fixed $f_{mw}$ = 10.4 GHz. The quantized voltage plateaus are clearly observed with the same voltage interval $\Delta V = 21.5$ μV, which is consistent with the theoretical value $hf_{mw}/2e$. The variation of $f_{mw}$ yields similar results, as shown in Fig. 3b, where $\Delta V$ is linearly proportional to $f_{mw}$ with the slope $h/2e$ = 2.07 μV/GHz (solid line in the inset). The current width of the $n$th Shapiro step $\Delta I_n$ are depicted in Fig. 3c as a function of $P^{1/2}$, which fit well with the theoretical expression[1] $\Delta I_n = 2I_C \left| J_n(2eV_{mw}/hf_{mw}) \right|$, where $J_n$ is the $n$th order Bessel function and $V_{mw}$ is the amplitude of the microwave voltage across the junction.

Additional evidence for the Josephson coupling in the PbS NWs can be found in the dependence of $I_C$ on the magnetic field $B$ perpendicular to the substrate. Figure 4a shows the color plot of the differential resistance $dV/dI$ as a function of $B$ and $I$, where the supercurrent region is denoted by the dark blue color. We note that $I_C$ decreases monotonically with increasing $B$ and vanishes at $B$ ~ 0.16 T. The $I_C(B)$ data (symbols) in Fig. 4b is quite different from the conventional expectation of a periodic modulation of $I_C$ with $B$, i.e., the Fraunhofer diffraction pattern,[1] where the $I_C$ minima occur at the integer magnetic flux quanta in the junction area. Our observations can be explained using the narrow junction model,[25] where the applied $B$ field acts as a pair breaker for the Cooper pairs induced in the normal conductor with a junction width $w$ smaller than or comparable to the magnetic length $\xi_B = (\Phi_0/B_0)^{1/2}$, where $\Phi_0 = h/2e$ is the magnetic-flux quantum and $B_0 = \Phi_0/Lw$ (see Supporting Information). Here, we obtained $\xi_B = 157$ nm for device **D1** with $w = 130$ nm, satisfying a narrow-junction condition of $w < \xi_B$. Similar features have also been observed in other narrow superconducting junctions.[13, 16, 22, 26] Additionally, the $I_C$ hump near zero field is attributed to the magnetic field focusing effect.[15]

Another peculiar feature of the $I_C(B)$ data is shown in Fig. 4c for device **D3** with $w = 200$ nm and $\xi_B = 190$ nm, resulting in $w \gtrsim \xi_B$. We note that there exist two different types of $I_C(B)$ patterns, both indicated by $dV/dI$ peak structures. The inner structure enclosing the dark blue region resembles a Fraunhofer-type modulation of $I_C$ with $B$, which is commonly observed in a wide Josephson junction.[18] It follows the Fraunhofer relation $I_C(B) = I_C(0)|\sin[\pi(\Phi/\Phi_0)]/(\Phi/\Phi_0)|$, where $\Phi$ is the magnetic flux through the superconducting junction. Then, the first minimum of $I_C$ is expected to be located at the magnetic field $B_1 = \Phi_0/[(L + 2\lambda)w]$, where $\lambda$ is the London penetration depth of the superconducting electrodes.[18] Because $B_1$ is found to be 19 mT, we obtain $\lambda = 180$ nm for the PbIn electrodes, which is comparable to the value obtained in our previous study.[6] The abrupt switching of $I_C(B)$ above $B_1$ is attributed to the penetration of magnetic vortices into the PbIn electrodes. The outer $dV/dI$ peak structure, occurring at $I_C^*$, reveals a monotonic suppression of $I_C^*$ by $B$, which is consistent with the narrow junction model with $r = 0.53$ (see Supporting Information). Thus, we conclude that the monotonic suppression and periodic modulation of $I_C$ with $B$ can occurr simultaneously in intermediate-width ($w \gtrsim \xi_B$) junction.

As displayed in Fig. 5a, the repetition of the current sweep reveals that there exists a stochastic distribution of $I_C$ switching from the superconducting to the resistive branches (see Methods for measurement details). It is well known that the $I_C$ distribution is closely related to the dynamics of the Josephson phase particle in the superconducting junction,[1] where the $I_C$ switching event corresponds to the escape of the phase particle from the local minima of the washboard potential $U(\varphi) = -E_{J0}[\cos(\varphi) + (I/I_{C0})\varphi]$, where $\varphi$ is the phase difference across the junction, and $I_{C0}$ and $E_{J0}$ ($= \hbar I_{C0}/2e$) are the fluctuation-free $I_C$ and Josephson coupling energy,[1] respectively (see Fig. 5b), with assuming a sinusoidal current-phase relation. Then, the escape is governed by the thermal activation[27] (TA), phase diffusion[28, 29] (PD), and macroscopic quantum tunneling[30] (MQT) processes, depending on relative strength of $E_{J0}$ over thermal fluctuations.

Figure 5c shows the $I_C$ distribution data measured at various temperatures, indicating that the sharp $I_C$ distribution obtained at high temperature ($T = 1.2$ K) becomes remarkably broadened at lower temperatures. As a result, the standard deviation of the $I_C$ distribution increases monotonically with decreasing temperature down to $T = 0.46$ K, implying that the PD process is the dominant switching current mechanism in this experiment,[31] where thermally-activated phase particles are repeatedly retrapped in the neighboring potential minima because of a strong dissipation during the escape. The fitting lines of the switching probability $P(I_C)$ in the PD model are consistent with the observed $I_C$ distribution data, as shown in Fig. 5c. The switching probability is related to the escape rate $\Gamma$ by $P(I_C) = [\Gamma(I_C)/(dI/dt)] \left\{ 1 - \int_0^{I_C} P(I')dI' \right\}$.[27] Here, $dI/dt$ is the sweep rate of the applied current. Figure 5d reveals that the behavior of the current dependence of $\Gamma$ changes below $T = 0.33$ K, indicating that there occurrs a crossover from the PD to the TA regimes as the temperature decreases. The MQT behavior, however, was not observed in our experimental range, in contrast to graphene-based superconducting junctions.[31, 32] It may be caused by an enhanced electron temperature caused by an incomplete filtering of the high-frequency noise.

**CONCLUSIONS**

In conclusion, we fabricated and characterized the electronic transport properties of nanohybrid superconducting junctions made of PbS NWs contacted with PbIn superconducting electrodes. We observed the highest values of the critical current and $I_C R_N$ product ever reported in semiconductor-NW-based superconducting junctions. Very strong superconducting proixmity effects in PbS NWs are attributed to the formation of highly transparent semiconductor–superconductor contacts with PbIn electrodes, which yield the relatively large superconducting gap energy. Both the microwave and magnetic field dependences of the $I$–$V$ characteristics of the PbS-NW superconducting junctions are in good agreement with theoretical expectations. Measurement and analysis of the stochastic distribution of the switching current reveals the underlying mechanism of escape dynamics of the phase particle in the NW-based superconducting junctions.

## METHODS

**Synthesis.** The PbS nanowires (NWs) were synthesized using the chemical vapor deposition method in a horizontal tube furnace (Lindberg Blue M). Lead chloride (PbCl$_2$, 99.999%, Alfa Aesar) and Sulfur (S, 99.9999%, Alfa Aesar) powders were placed in the center and outside the heating zone of the furnace, respectively. The growth substrate was prepared using electron-beam evaporation of a 100 nm Ti thin film onto a SiO$_2$ coated Si wafer. The substrate was placed 5 cm downstream from the center of the heating zone ($T \sim 550°C$). The synthesis system was first evacuated to a base pressure of 15 mTorr, and then flushed with N$_2$ (99.999%) three times before filling to atmosphere pressure. The furnace temperature was quickly ramped to 630°C at 60°C/min, while 150 sccm N$_2$ flow was maintained. At the peak temperature, the quartz boat containing S was transferred to the heating zone to trigger the growth. The growth duration varied from 30 min to 2 h. After the growth, the furnace was naturally cooled down to room temperature over approximately 3 h. The N$_2$ flow was kept during the entire cooling down process to remove any S residue. The doping of the PbS NWs was modulated by varying the weight ratio of $PbCl_2$ and S. The more details of the synthesis, characterization, and doping level control can be found in a previous publication.[17]

**Device Fabrication and Transport Measurements.** For the fabrication of the PbS-NW-based superconducting junctions, the as-grown PbS NWs were transferred onto a highly doped $n$-type silicon substrate covered by a 300-nm-thick oxide layer with a prepatterned bonding pad. The superconducting electrodes were patterned using electron-beam lithography. Prior to the metal evaporation, the NWs were exposed to oxygen plasma to remove any residual resist. To make transparent interfaces, the native oxide layer on the PbS NW surface was etched in 6:1 buffered oxide etchant for 10 s. The superconducting electrodes with $Pb_{0.5}In_{0.5}$ (250 nm)/Au (10 nm) were deposited using electron-beam evaporation. A highly doped $n$-type silicon substrate served as the back-gate electrode. Figure 1b shows the scanning electron microscope (SEM) image of a typical PbS-NW-based Josephson junction with a channel length $L$ = 190 nm and the four-probe measurement configuration. We also measured the transport properties of 14 devices in three different refrigerator systems: [3]He refrigerator (cryogenic) with a base temperature of $T$ = 0.3 K, a 1.5 K cryostat using microwave response, and a 2.5 K cryostat. In particular, the switching current was observed in devices with channel lengths that were less than 250 nm. The detailed parameters of the devices are shown in Table S1. To obtain the switching current distribution, a triangle-wave-shape bias current was applied to the sample with a ramping rate $dI/dt$ = 240 $\mu As^{-1}$, while 5,000 $I_C$ data were recorded at a threshold voltage $V_{th}$ = 30 $\mu V$ and a fixed temperature.

**ASSOCIATED CONTENT**

**Supporting Information**. The supporting information is available free of charge on the ACS

Publications website at http://pubs.acs.org.

**AUTHOR INFORMATION**

**Corresponding Author**

*E-mail: yjdoh@gist.ac.kr

**Notes**

The authors declare no competing financial interest.

**ACKNOWLEDGMENT**

We are grateful to Dr. M.-H. Bae for useful discussions. This work was supported by the U.S. National Science Foundation (Grant No. DMR-1310678) and the National Research Foundation of Korea through the Basic Science Research Program (Grant No. 2015R1A2A2A01006833 and 2015R1A6A3A01020240) and the SRC Center for Quantum Coherence in Condensed Matter (Grant No. 2016R1A5A1008184).

**TOC Image:**

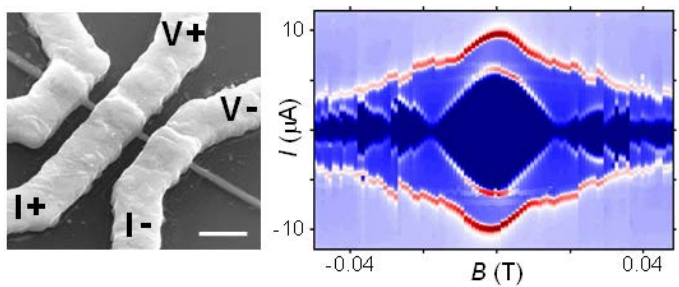

**Figure Legends**

**Table 1.** Detailed properties of the PbS NW devices and previously reported nanowire superconducting junctions

**Figure 1.** SEM images of (a) as-synthesized PbS NWs, and (b) PbS-NW-based junction contacted with PbIn superconducting electrodes. In a four-probe measurement setup, the current bias was driven between I+ and I−. Meanwhile, the voltage drop was measured between V+ and V−. (c) Resistance vs. temperature curve plotted for device **D3**. Superconducting transition temperature of PbIn is $T_{\mathrm{C,SC}} = 7.0$ K, while the junction resistance is fully suppressed below $T_{\mathrm{C,JJ}}$ = 3.5 K. (d) $I$–$V$ characteristics of devices **D1** and **D2** at $T = 0.3$ K. Here, $I_{\mathrm{C}}$ ($I_{\mathrm{R}}$) means the switching (retrapping) current. (e) Differential conductance vs. voltage curve for device **D1**. The arrows indicate the subgap conductance peaks (see text). (f) Gate-voltage dependence of $I_{\mathrm{C}}$ and $R_{\mathrm{N}}$ for device **D4**.

**Figure 2.** (a) Temperature dependence of the $I$–$V$ characteristic curves for device **D2**. Plots are offset for clarity. (b) Temperature dependences of $I_{\mathrm{C}}$ (circles) and $I_{\mathrm{R}}$ (squares). The solid line is a theoretical calculation of $I_{\mathrm{C}}$ using the long and diffusive junction model (see text).

**Figure 3.** (a) Microwave-power dependence of the $I$–$V$ characteristics under the external microwave at the frequency $f_{\mathrm{mw}} = 10.4$ GHz. The measurement temperature was $T = 1.5$ K. Voltage plateaus (or Shapiro steps) are clearly observed at $V = V_{\mathrm{n}}$ with $n = 1, 2, 3$. (b) $I$–$V$ characteristics at $f_{\mathrm{mw}} = 4.4, 7.3, 10.4,$ and 15.5 GHz. Here, the voltage axis was normalized by $hf_{\mathrm{mw}}/2e$. The inset shows the frequency dependences of the voltage interval, $\Delta V$, data (squares)

and the theoretical prediction (solid line) based on the ac Josephson relation. (c) Current width $\Delta I_n$ of the $n$th Shapiro step ($n = 0, 1$) with respect to the square root of the microwave power ($P^{1/2}$). The solid lines correspond to the theoretical calculations (see text).

**Figure 4.** (a) Color plot of dynamic resistance $dV/dI$ as a function of $I$ and $B$ at $T = 0.3$ K. The dark blue region corresponds to the supercurrent flow region. (b) Magnetic-field dependence of the $I_C$ data (symbols). The solid line is a fit using the narrow junction model (see text). (c) Color plot of $dV/dI$ for device **D3** at $T = 1.6$ K. (d) The $I_C(B)$ data (red symbols) and $I_C^*$ values corresponding to the first (circles) and secondary (squares) $dV/dI$ peaks, respectively. The red solid line indicates the Fraunhofer pattern in the wide Josephson junction, while the black line represents the fitting result using the narrow junction model.

**Figure 5.** (a) Repetitive measurement of the stochastic switching current (black lines) and corresponding $I_C$ distribution (red). (b) Schematic diagram of the tilted-washboard potential and the escape processes of a phase particle confined in the potential well. Temperature dependence of the (c) switching current distribution and (d) escaping rate. The solid lines in (c) and (d) are from the theoretical fits to the TA ($T = 0.3$–$0.33$ K) and PD ($T = 0.46$–$1.2$ K) models, respectively.

**Table 1.**

| Sample | $I_C$ (μA) | $J_C$ (kA/cm$^2$) | $I_C R_N$ (mV) | $e I_C R_N / \Delta$ |
|---|---|---|---|---|
| **Pb-based InAs NW[15]** | 0.61 | 31 | 0.35 | 0.26 |
| **Al-based InAs NW[13]** | 0.8 | 16 | 0.13 | 0.98 |
| **Nb-based InN NW[22]** | 5.7 | 50 | 0.44 | 0.31 |
| **This work (D1)** | 15 | 110 | 0.29 | 0.36 |
| **This work (D2)** | 7.3 | 55 | 0.84 | 1.05 |

**Figure 1.**

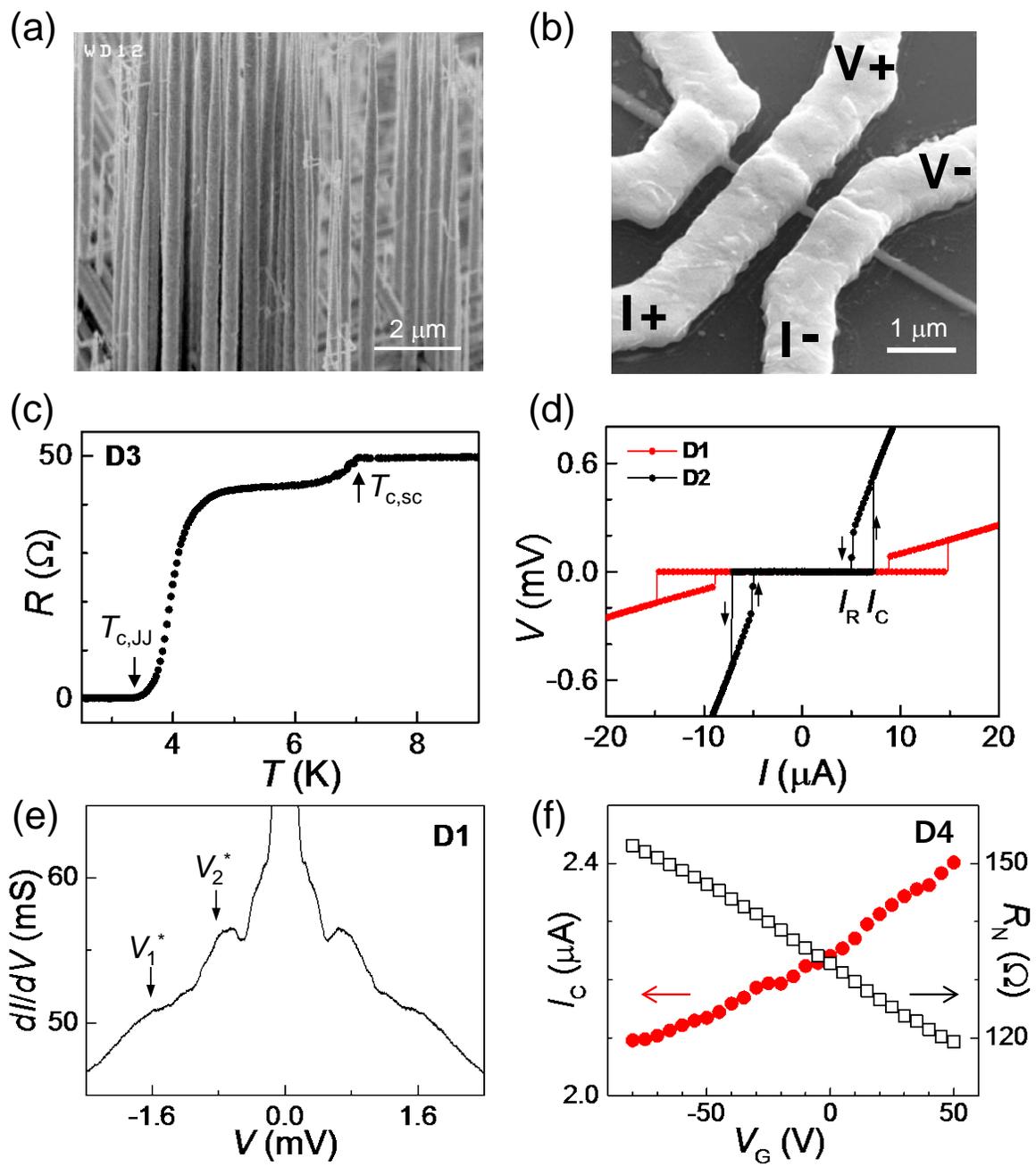

**Figure 2.**

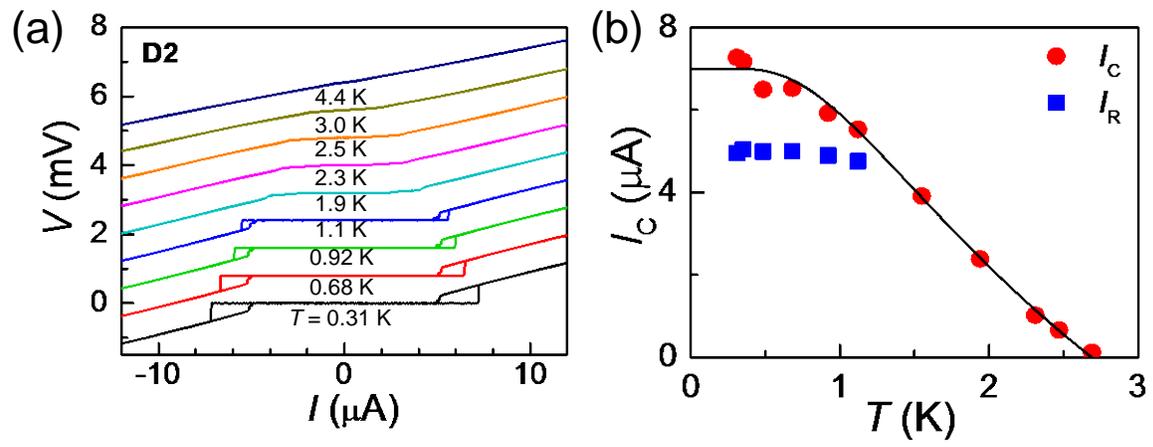

**Figure 3.**

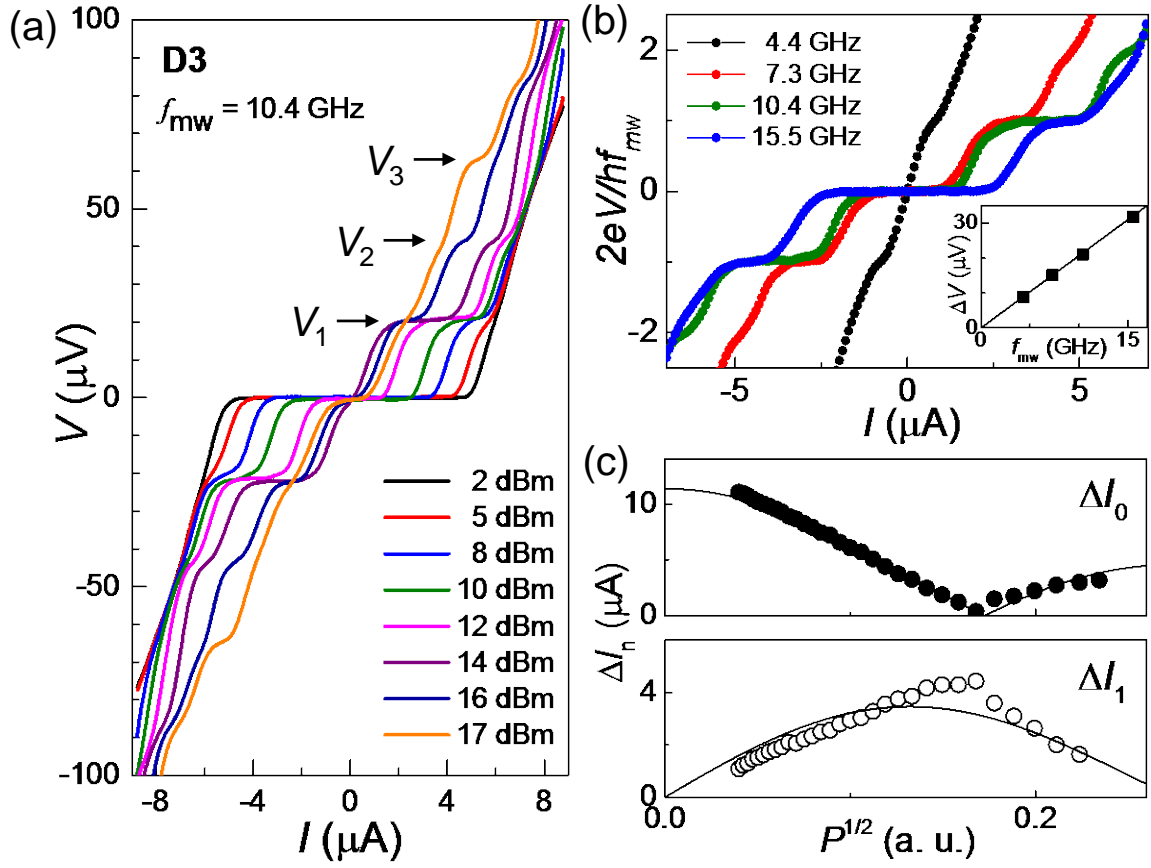

**Figure 4.**

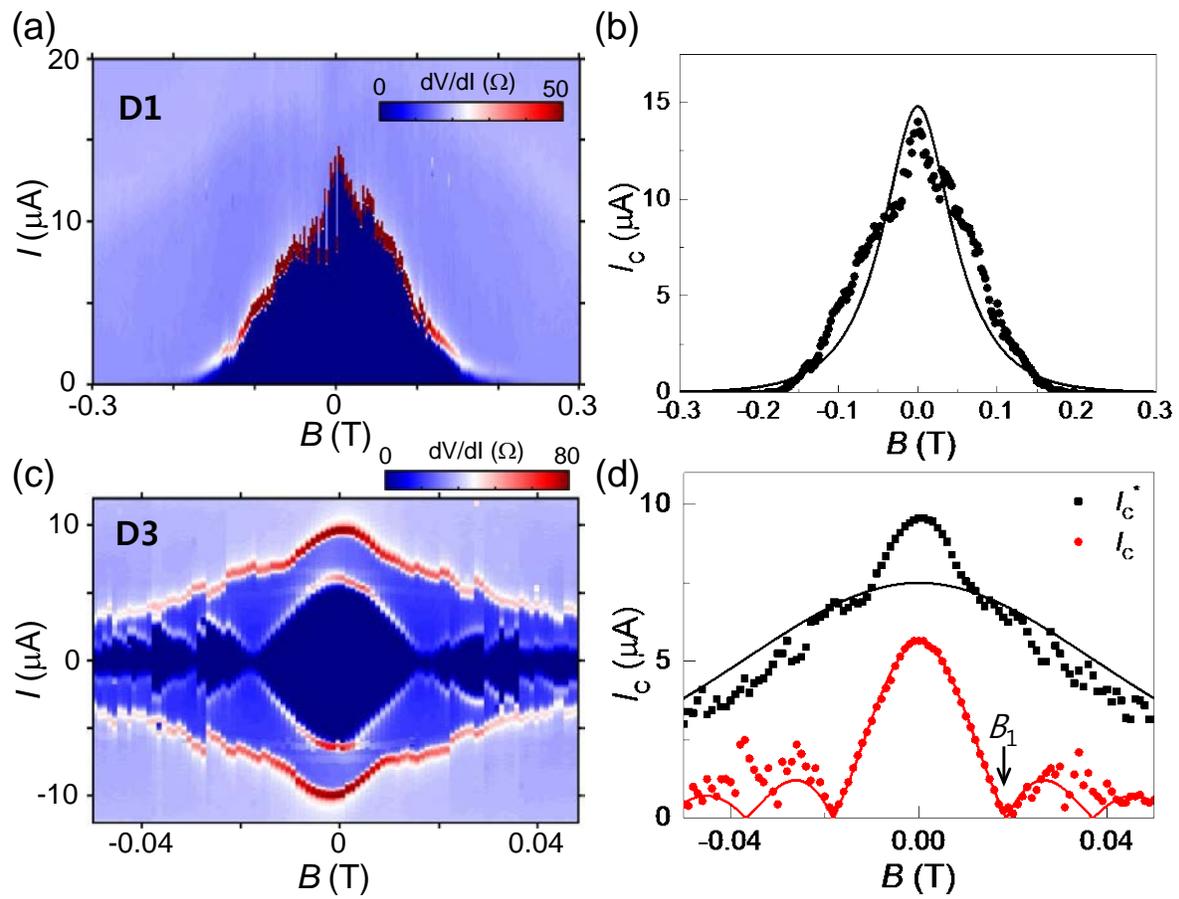

**Figure 5.**

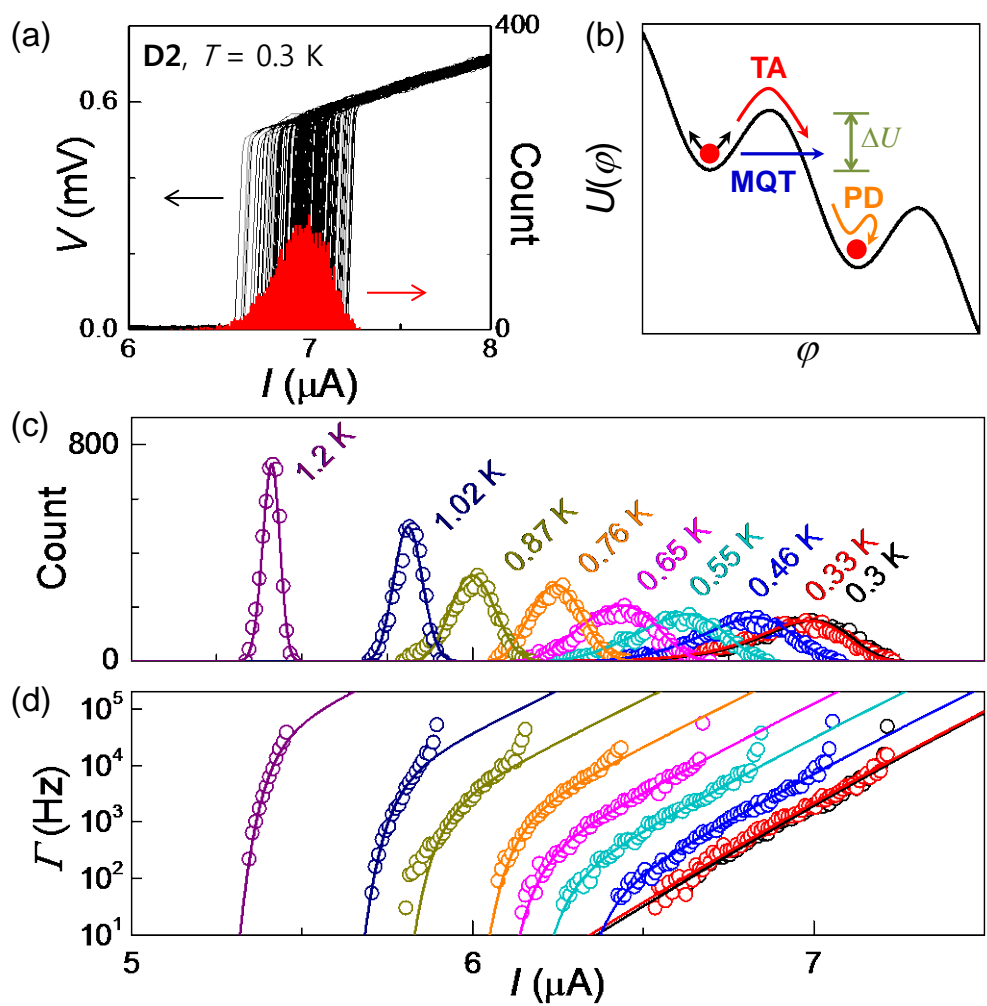

# Very Strong Superconducting Proximity Effects in PbS Semiconductor Nanowires

*Bum-Kyu Kim[1], Hong-Seok Kim[1], Yiming Yang[2], Xingyue Peng[2], Dong Yu[2], Yong-Joo Doh[1,*]*

[1]Department of Physics and Photon Science, Gwangju Institute of Science and Technology (GIST), Gwangju, 61005, Korea

[2]Department of Physics, University of California, Davis, CA 95616, USA

## Supporting Information

*Address correspondence to yjdoh@gist.ac.kr

**Table S1.** Physical parameters of PbS nanowire samples.

| No. | Channel length (nm) | Width (nm) | $I_C$ (μA) | Base temperature (K) | $R_N$ (Ω) |
|---|---|---|---|---|---|
| **D1** | 190 | 130 | 15 | 0.3 | 21 |
| **D2** | 190 | 130 | 7.3 | 0.3 | 120 |
| **D3** | 180 | 200 | 5.9 | 1.5 | 49 |
| **D4** | 185 | 125 | 3.2 | 2.6 | 130 |
| **D5** | 3000 | 140 | 0 | 0.3 | 133 |

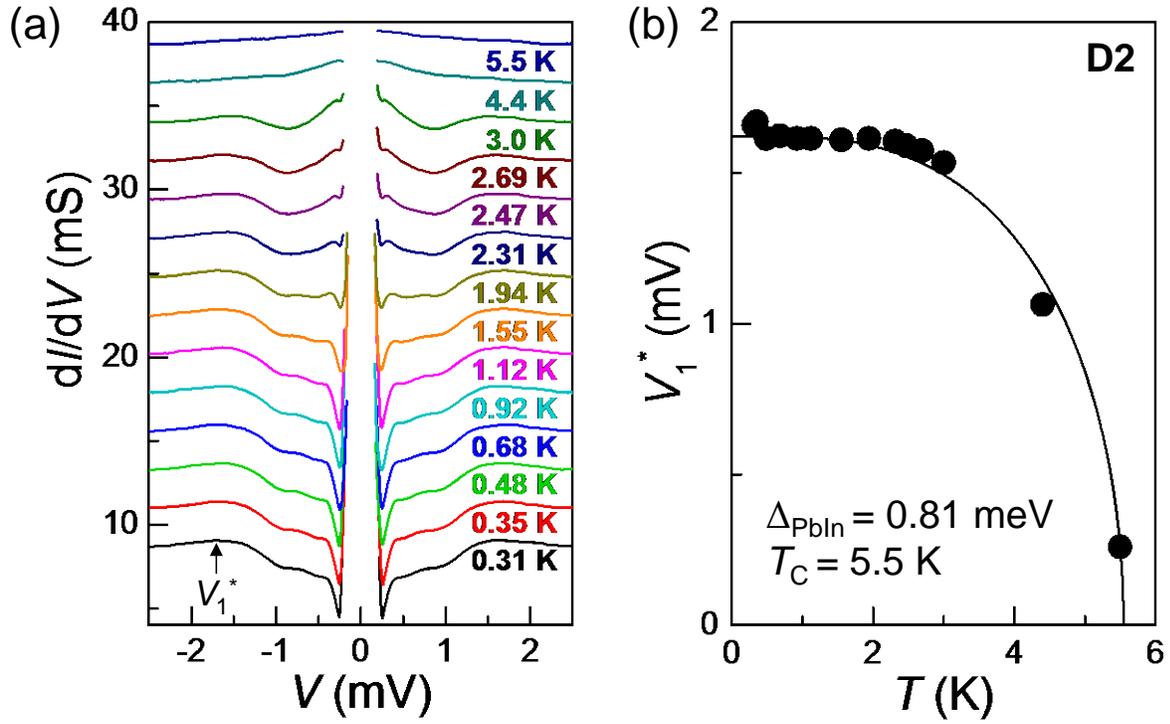

**Figure S1.** (a) Temperature dependence of *dI/dV* vs. *V* curve. Overshoot behavior near zero bias voltage is not shown here for clarity. (b) Temperature dependence of $V_1^*$ (symbols), corresponding to $\Delta_{PbIn}/e$. The solid line is a fit to the BCS theory.

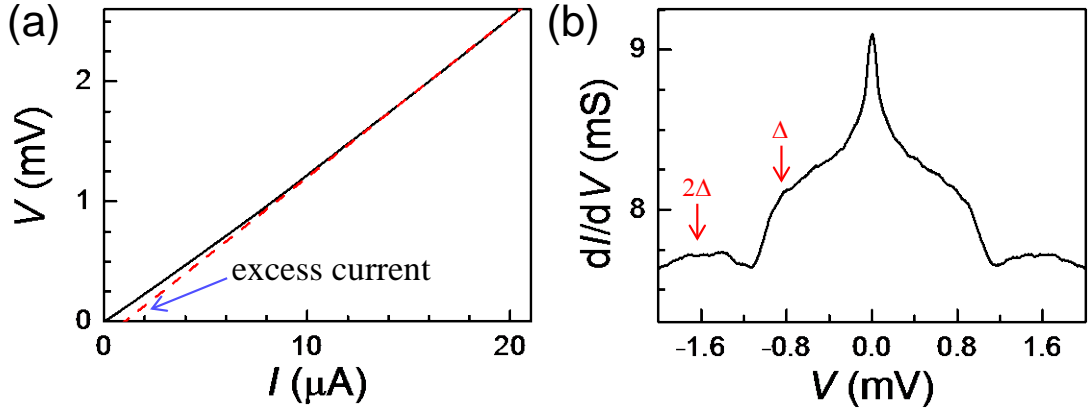

**Figure S2.** (a) *I-V* characteristic of PbS NW-based superconducting junction (**D5**) at $T = 300$ mK. The excess current, which is the zero-voltage crossing point of the extrapolated line of the *I-V* curve at the high bias ($eV > 2\Delta_{PbIn}$) region, is obtained to be $I_{exc} = 1$ μA. The transparency at the interface is estimated to be $T_{int} = 0.86$.[1] (b) Corresponding differential conductance *dI/dV* vs. *V* plot. Multiple Andreev reflections result in the dI/dV peaks occurring at $V^* = \Delta_{PbIn}/e$ and $2\Delta_{PbIn}/e$. Note that the conductance enhancement was observed from a very long channel ($L = 3$ μm) device, although no supercurrent was observed in **D5**.

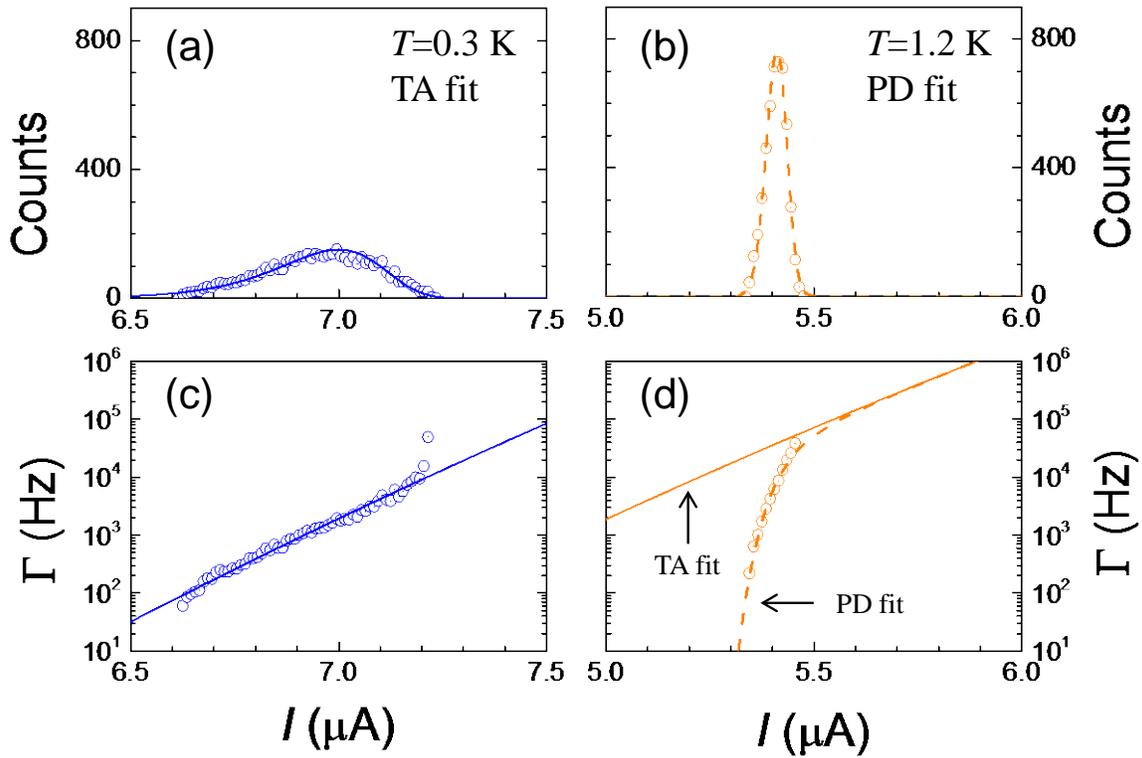

**Figure S3.** Fitting to the TA and PD models. Experimental data (symbols) of the switching current distribution and best-fit results (lines) at (a) $T = 0.3$ K and (b) 1.2 K, respectively. In the TA regime at $T = 0.3$ K, $P(I_C)$ shows a characteristic tail feature at the left side, while $P(I_C)$ in the PD regime exhibits a relatively symmetric and sharper distribution. (c-d) Corresponding escape rate $\Gamma(I_C)$ data (symbos) and best-fit results (solid lines for TA and dashed for PD model) at $T = 0.3$ K and 1.2 K, respectively.

**Fitting to the narrow junction model**

Using this model,[2] the monotonically decreasing $I_C$ can be expressed as

$$eRI_C = \frac{4\pi k_B T}{r} \sum_{n=0}^{\infty} \frac{\Delta^2/(\Delta^2+\omega_n^2)}{\sqrt{2\left(\frac{\omega_n+\Gamma_B}{E_{Th}}\right)}\sinh\sqrt{2\left(\frac{\omega_n+\Gamma_B}{E_{Th}}\right)}},$$

where $\Gamma_B = De^2B^2w^2/6\hbar$ is the magnetic depairing energy, $\omega_n = \pi k_B T(2n+1)$ is the $n$th Matsubara energy, $r = R_B/R_N$, and $R_B$ is the barrier resistance. The black solid line in Fig. 4b (Fig. 4d) is the fitting result with $r = 1.43$ (0.53), while $\Delta_{PbIn} = 0.81$ meV and $D = 103$ cm$^2$/s are fixed.

**Fitting to switching current models**

In TA regime,[3] switching event from superconducting to resistive branch occur thermally activated escape of phase particle in a tilted-washboard potential. The escape rate $\Gamma_{TA}$ is expressed by $\Gamma_{TA} = a_t(\omega_p/2\pi)\exp[-\Delta U/k_B T]$, where damping-dependent factor $a_t = (1+1/4Q^2)^{1/2}-1/2Q$, tilted washboard potential barrier $\Delta U = 2E_{J0}[(1-\gamma^2)^{1/2}-\gamma\cos^{-1}\gamma]$, normalized current $\gamma = I/I_{C0}$, Josephson plasma frequency $\omega_p = \omega_{p0}(1-\gamma^2)^{1/4}$, the plasma frequency in zero bias current $\omega_{p0} = (2eI_{C0}/\hbar C)^{1/2}$, and quality factor $Q = 4I_C/\pi I_R$. In this work, the geometrical junction capacitance $C$ was replaced by the effective capacitance $C_{eff} = \tau/R_N$, where electron diffusion time $\tau = \hbar/E_{Th}$, Thouless energy $E_{Th} = \hbar D/L^2$, diffusion constant $D = v_F l/3$, the Fermi velocity $v_F$, and the mean free path $l$.

In the PD regime,[4, 5] a thermally activated phase particle to escape at low bias current regime is repeatedly retrapped in the neighboring potential well due to a strong dissipation. Thus the escape rate $\Gamma_{PD}$ is given[4] by $\Gamma_{PD} = \Gamma_{TA}(1-P_{RT})\ln(1-P_{RT})^{-1}/P_{RT}$. The retrapping probability $P_{RT}$ is obtained by integrating the retrapping rate $\Gamma_{RT} = \omega_{p0}[(I-I_{R0})/I_{C0}](E_{J0}/2\pi k_B T)^{1/2}\exp(-\Delta U_{RT}/k_B T)$, where $\Delta U_{RT} = (E_{J0}Q_0^2/2)[(I-I_{R0})/I_{C0}]^2$ is the retrapping barrier with $Q_0 = 4I_{C0}/\pi I_{R0}$ and $I_{R0}$ is the fluctuation-free retrapping current.[6, 7]

**Supplementary Reference**